\documentclass[12pt,twoside]{article}   
\usepackage[super,sort,comma]{natbib}

\usepackage{fancyhdr}		

\usepackage[section]{placeins}   %

\usepackage{graphicx,float,caption,subcaption}
\usepackage{amsmath,amssymb,amsfonts,extarrows}
\usepackage{algorithmicx}
\usepackage{textcomp}
\usepackage{algorithm}
\usepackage{algpseudocode}

\DeclareMathOperator*{\argmin}{arg\,min}

\makeatletter \renewcommand\@biblabel[1]{$^{#1}$} \makeatother
\newcommand{\note}[1]{\mbox{}\\ \noindent \rule{16cm}{0.5mm} \\
{\em #1} \\ \noindent \rule{16cm}{0.5mm}
\typeout{    }
\typeout{***********note active on this page *************************}
\typeout{Note: #1  }
\typeout{****************************************end Note}
}

\renewcommand{\note}[1]{}

\newcommand{\cen}[1]{\begin{center} #1 \end{center}}

\usepackage[mathlines]{lineno}

\usepackage{hyperref}
\hypersetup{ colorlinks,
    citecolor=blue,
    filecolor=blue,
    linkcolor=blue,
    urlcolor=blue
}

\usepackage{xcolor}

\definecolor{gray}{rgb}{0.6,0.6,0.6}
\definecolor{red}{rgb}{0.85,0,0}
\definecolor{green}{rgb}{0,0.85,0}
\definecolor{blue}{rgb}{0,0,0.85}
\definecolor{beige}{rgb}{0.92,0.87,0.78}
\usepackage[all]{hypcap}    

\graphicspath{{figures/}}

\def\R{\mathbb{R}}

\def\rR{\mathcal{R}}

\def\s{\sigma}
\def\S{\Sigma}

\begin{document}

\cen{\sf {\Large {\bfseries Optimizing dual-energy CT technique for iodine-based contrast-to-noise ratio} \\  
\vspace*{5mm}
Fatma Terzioglu$^1$, Emil Y. Sidky$^2$, Jp Phillips$^2$, Ingrid S. Reiser$^2$, Guillaume Bal$^3$, and Xiaochuan Pan$^2$
 } \\
$^1$Department of Mathematics, North Carolina State University, 2311 Stinson Dr.,  Raleigh, NC 27695, USA\\
$^3$Departments of Statistics and Mathematics, The University of Chicago, 5747 S. Ellis Ave., Chicago, IL 60637, USA\\
$^2$Department of Radiology, The University of Chicago, 5841 S. Maryland Ave., Chicago, IL 60637, USA
\vspace{3mm}\\
}

\pagenumbering{roman}
\setcounter{page}{1}
\pagestyle{plain}
Author to whom correspondence should be addressed. email: fterzioglu@ncsu.edu\\

\begin{abstract}
\noindent
{\bf Purpose:}
The goal of this study is to propose a systematic method for determining the optimal x-ray tube settings/energy windows and fluence for minimal noise and maximum CNR in material density images obtained from dual-energy CT (DECT) scans by fixing the subject size and the total radiation dose.\\
{\bf Methods:}
The noise propagation in the process of sinogram and image reconstruction from DECT measurements is analyzed. The main objects of the study are the pixel variances for the sinogram and monochromatic image and the contrast-to-noise ratio (CNR), which were shown to depend on the Jacobian matrix of the sinograms-to-DECT measurements map.

Analytic estimates for the sinogram and monochromatic image pixel variances and the CNR as functions of tube potentials, fluence, and virtual monochromatic image (VMI) energy are derived, and then used in a phantom experiment as an objective function for optimizing the tube settings to minimize the image noise and maximize the CNR.\\
{\bf Results:} 
A non-trivial example that shows the existence of singular solutions to the inversion of sinograms-to-DECT measurements map was presented. Additionally, the optimal VMI energy for maximal CNR was determined. The optimal energy VMI was found to be the least noisy monochromatic image synthesized from the iodine and water density images, and it was shown that using more general weights in combining the two images linearly does not improve image quality. 

When the x-ray beam filter material was fixed at 2mm of Aluminum and the photon fluence for low and high kV scans were considered equal, the tube potential pair of 60/120 kV led to the maximal CNR in the VMI formed at energy 55 KeV. \\
{\bf Conclusions:} 
Optimizing DECT scan parameters to maximize the CNR can be done in a systematic way. Also choosing the parameters that maximize the Jacobian determinant over the sinogram domain would lead to more stable reconstructions due to the reduced amplification of the measurement noise. Since the values of the Jacobian determinant depend strongly on the imaging task, careful consideration of all of the relevant factors is needed when implementing the proposed framework.\\
\end{abstract}


\newpage     

\tableofcontents

\newpage

\setlength{\baselineskip}{0.7cm}      

\pagenumbering{arabic}
\setcounter{page}{1}
\pagestyle{fancy}

\section{Introduction}\label{sec:introduction}
Dual-energy CT (DECT) systems enable the simultaneous acquisition of two spectral measurements to identify different materials within the scanned object. DECT has been demonstrated to outperform single-energy CT in terms of image quality and contrast-to-noise ratio (CNR), allowing for reduced radiation exposure and contrast agent concentration while maintaining image quality \cite{Grant2014, Leng2015, Kalisz2018, Sakabe2018, Tabari2020}.

The concept of DECT was introduced by Hounsfield in the 1970s \cite{Hounsfield}, and the mathematical framework for pre-reconstruction processing of DECT data was developed by Alvarez and Macovski in a seminal paper in 1976 \cite{AlvarezMacovski}. Their approach to material decomposition is based on the assumption that energy-dependent attenuation coefficients of chemical compounds can be approximated by a linear combination of elemental mass attenuation maps weighted by the partial density of each element. This reduces sinogram reconstruction to solving a nonlinear system of equations for each sinogram value.

Due to the nonlinearity of the DECT measurement model, the uniqueness of reconstructed sinograms from DECT measurements is not guaranteed. Levine first provided an example of non-unique solutions in DECT for a material basis of water and bone using spectra with three discrete photon energies \cite{Levine}. Alvarez analyzed the non-invertibility of the sinogram to DECT measurements mapping by studying the Jacobian determinant \cite{Alvarez2019}. In general, the nonvanishing of the Jacobian determinant only guarantees local uniqueness (a.k.a. injectivity), requiring additional constraints for global uniqueness. Bal and Terzioglu\cite{BalTer20} presented sufficient analytic criteria for the global injectivity of multi-energy CT (MECT) measurement map for a given number of materials and equal number of energy measurements. In the case of DECT, they showed that the nonvanishing of the Jacobian determinant of the sinogram to DECT measurements map throughout its domain is sufficient to ensure global uniqueness. They also demonstrated how the choice of basis materials and x-ray spectra influences the Jacobian determinant values and, consequently, the invertibility. In this paper, we showcase the occurrence of nonuniqueness when the Jacobian determinant vanishes within the rectangular region encompassing all possible sinogram values, providing a clear example of two isolated solutions corresponding to the same DECT measurement pair.

It was shown by Bal et al. \cite{BalGongTer} that the stability of the inversion of the mapping of sinograms to MECT measurements is improved by choosing the x-ray spectra that maximize Jacobian determinant values of the sinogram to DECT measurements map. In this paper, we propose a systematic method for determining optimal tube settings and fluence that minimize noise and maximize CNR in material density images obtained from DECT scans while keeping subject size and total radiation dose fixed. To achieve this, we consider a noise model based on a compound Poisson process for the DECT measurements and analyze the noise propagation from DECT measurements to material density images by linearizing the inverse sinogram-to-photon-counts transformation. We derive analytic expressions for sinogram and monochromatic image pixel variances as functions of tube potentials, fluence, and virtual monochromatic image (VMI) energy. These expressions, along with CNR, are used as an objective function to optimize the tube settings and minimize image noise in a phantom experiment. We determine the optimal VMI energy for maximal CNR and also prove that the least noisy monochromatic image synthesized from iodine and water density images corresponds to the optimal energy VMI. Consequently, using more general weights to linearly combine the two images does not enhance image quality.

The problem of improving iodine CNR in VMI has been previously addressed, but our approach differs from existing works in terms of the analysis methods employed. Specifically, the works by Leng et al. \cite{Leng2015} and Tao et al. \cite{Tao2019} focus on applying denoising techniques during the reconstruction process. Yu et al. \cite{Yu2009, Yu2011} examine the effect of subject size and photon fluence on the image quality of linearly mixed images generated from DECT scans using a dual-source CT scanner, assuming a fixed x-ray source kilovoltage-peak (kV) setting and total radiation dose. Michalak et al. \cite{Michalak2017} conduct a phantom study to empirically determine the optimal photon energies for virtual mono-energetic imaging across various phantom sizes. Dabli et al. \cite{Dabli2021} empirically determine optimal tube potential settings that yield high image quality and accuracy for low iodine concentration quantification. Ren et al. \cite{Ren2021} analyze the conditioning of spectral weights by employing singular value decomposition of the matrix formed by sampling the intensity profile of each spectral weight over the energy range.

\section{Methods}\label{sec:methods}
In this section, we detail the approaches and procedures employed to optimize Dual-Energy Computed Tomography (DECT) scan parameters for improved iodine-based Contrast-to-Noise Ratio (CNR). We first present the physical model for the DECT measurements, considering the well-established assumption of the basis material decomposition \cite{AlvarezMacovski, Williamson2006} and a noise model based on the compound Poisson process \cite{Whiting2002}. Next, we explore the role of the Jacobian matrix in ensuring reconstruction uniqueness and examine noise propagation by linearizing the inverse sinogram-to-photon-counts transformation. We then derive analytical expressions for CNR, mean sinogram variance, and mean pixel variance. We finally show that the smallest eigenvalue of the mean pixel covariance matrix gives the minimum mean variance for VMI obtained by using optimal VMI energy, identifying optimal energy levels for maximum image quality.

\subsection{Scan configuration, DECT technique, and noise simulation}
The physics modeling for the dual-energy CT system includes available models for Tungsten X-ray source spectra, low and high kV fluence, response of energy-integrating detectors, and compound Poisson noise \cite{Whiting2002}. 

For iodine-based contrast imaging with a DECT system, we assume that the scanned object is composed of iodine and water only. Consequently, the linear attenuation coefficient is approximated by\cite{AlvarezMacovski}
\begin{align}
    \mu(E,y) \approx M_1(E)\rho_1(y)+M_2(E)\rho_2(y),
\end{align}
where $M_1(E)$ and $M_2(E)$ denote material attenuation at energy $E$ for iodine and water, respectively, and $\rho_1(y)$ and $\rho_2(y)$ are their mass density at a spatial location $y$. While the mass densities are unknown and need to be reconstructed, the energy dependent attenuation maps are known \emph{a priori}, which are available at the NIST database\cite{NIST}. To simplify the notation, we write
\[
M(E) =
\begin{bmatrix}
M_1(E)\\
M_2(E)
\end{bmatrix}.
\]
For a given x-ray beam $l$, the x-ray transform (or sinogram) of mass density $\rho_j$ of the $j$-th material is given by $x_j(l)=\textstyle \int_{l} \rho_j(y)dy$. We let 
\[
x=x_l =
\begin{bmatrix}
x_1(l)\\
x_2(l)
\end{bmatrix}.
\]
Let $S_1(E)$ and $S_2(E)$ be the x-ray energy spectra corresponding to low and high energy x-ray tube potentials $tp_1$ and $tp_2$, respectively. The x-ray spectra used in the experiments were known \emph{a priori}, which were modeled, for given tube potentials, by using the publicly available Python software toolkit SpekPy v2.0\cite{Spekpy2}.

For an x-ray beam $l$, the number of photons incident on the detector with energy $E$ per unit time corresponding to the $i$-th measurement is given by
\begin{align}\label{eq:IiE}
     I_i(x_l;E) =  S_i(E)e^{-\int_l \mu(E,y)dy} \approx  S_i(E) e^{-M(E) \cdot x_l}, \quad 1\leq i\leq 2,
\end{align}
where $M(E) \cdot x_l = M_1(E)x_1(l)+M_2(E)x_2(l)$.

 The total number of detected photons associated to line $l$ is then given by
\begin{align}\label{eq:Ii}
  I_i(x_l) = \int_0^\infty S_i(E) e^{-M(E) \cdot x_l} D(E) dE, \quad 1\leq i\leq 2,
\end{align}
where $D(E)$ is the energy dependence of the detector sensitivity. For energy integrating detectors, $D(E)=\alpha E$ for some $\alpha>0$. \cite{Whiting2002}

In this study, we consider the negative logarithm of the intensity measurements
\begin{equation}\label{eq:g}
  g_i(x) = - \ln I_i(x), \quad i=1,2.
\end{equation}

We define 
\[
I(x) =
\begin{bmatrix}
I_1(x)\\
I_2(x)
\end{bmatrix}, \qquad
g(x) =
\begin{bmatrix}
g_1(x)\\
g_2(x)
\end{bmatrix}.
\]

We assume that the sinograms $x=x_l\in \rR=[0,a_1] \times [0,a_2]$, a rectangle in $\R^2$ with $a_j$ being the maximal attenuation over all possible lines expected for the $j$-th material in a given imaging task.

In the reconstruction of mass density maps from DECT measurements, we consider a two step method given in the following diagram. For each line $l$,
\begin{align}
\begin{bmatrix}
 g_1\\ 
 g_2
\end{bmatrix}
\quad \xlongrightarrow[\text{method}]{\text{Newton's}} \quad
\begin{bmatrix}
 x_1\\
 x_2
\end{bmatrix}
\quad \xlongrightarrow[\text{back-projection}]{\text{Filtered}} \quad
\begin{bmatrix}
 \rho_1\\
 \rho_2
\end{bmatrix}.
\end{align}

For $i=1,2$, the spectral measurements $I_i$ are assumed to be independent random variables that follow a compound Poisson process \cite{Whiting2002}. The covariance matrix for the log-intensity measurements $g$ is then given by
\begin{align}
\Sigma_g(x) = 
\begin{bmatrix}
\s_g(x)_1^2 & 0\\
0 & \s_g(x)_2^2
\end{bmatrix},
\end{align}
where 
\begin{align}\label{noise}
  \s_g(x)_i^2 = \displaystyle \frac{\int_0^\infty  D^2(E) S_i(E) e^{-M(E)\cdot x}dE}{\Big(\int_0^\infty  D(E) S_i(E) e^{-M(E)\cdot x}dE\Big)^2},
\end{align}
is the variance of the $i$-th log-intensity measurement $g_i$. 

We note that $\Sigma_g$ depends on the detector sensitivity $D(E)$ but is independent of the factor $\alpha$ if $D(E)$ is replaced by $\alpha D(E)$.

\subsection{Jacobian of the sinogram-to-photon-counts transformation and uniqueness of reconstructions}\label{subsec:Jacobian}
 Unlike the case of single energy CT, the reconstructions obtained from DECT measurements may not always be unique, which is mainly due to the non-linearity of the DECT measurement model with respect to sinogram values. For nonlinear maps defined on convex domains, local constraints on the Jacobian of the forward measurements provide sufficient criteria for the uniqueness of reconstructions \cite{GaleNikaido, Garcia, MasColell}.
 
 In our case, the map $x\to g(x)$ is smooth and its Jacobian matrix at point $x\in \rR$ is given by the matrix $J(x)$ with entries
\begin{align}
  J_{ij}(x) = \frac{\partial g_i}{\partial x_j}(x) = \frac{\int_0^\infty D(E)S_i(E) M_j(E) e^{-M(E)\cdot x} dE}{\int_0^\infty D(E)S_i(E) e^{-M(E)\cdot x} dE} ,\qquad i,j =1,2.
\end{align}

 Based on the work of Gale and Nikaido\cite{GaleNikaido}, Bal and Terzioglu \cite{BalTer20} proved that if the Jacobian determinant 
\begin{align}\label{detJ}
\det J(x) = J_{11}(x)J_{22}(x)-J_{12}(x)J_{21}(x) \neq 0,
\end{align}
for all $x \in \rR$, the map $x\to g(x)$ is globally injective. This means that the reconstruction of $x$ from the knowledge of $g(x)$ (or $I(x)$) is unique. 

In DECT, the values of the Jacobian determinant depend on the material basis and the x-ray spectra \cite{BalTer20}. For iodine-water material pair, we demonstrate the dependence of the Jacobian determinant on the x-ray spectra in Figures \ref{fig:signedDetJ} and \ref{fig:detJplot} of section \ref{sec:uniqueness} .

We also present in Fig. \ref{fig:nonUniqueness} an example scan protocol where the uniqueness does not hold. That is, the Jacobian determinant vanishes inside the rectangle and there exist two distinct sinogram values that are mapped by $g$ to the same measurement pair.

\subsection{Noise propagation based on linearization of the inverse sinogram-to-photon-counts transformation}
We now present an analysis of the noise propagation from DECT measurements to the reconstructed sinograms by considering first order Taylor approximation to inverse sinogram-to-photon-counts transformation. 

For a given sinogram value $x_l \in \rR$ that corresponds to a line $l$, we let $g^\eta(x_l)$ denote the noisy DECT measurement:
\begin{align}\label{eq:geta}
    g^\eta(x_l) = g(x_l) + \eta,
\end{align}
where $\eta$ is the noise vector. Let $x_l^{\eta}$ be the sinogram value that is reconstructed from $g^\eta(x_l)$. Assuming that the noise is small and considering a first order Taylor expansion, 
we have
\begin{align}\label{eq:errorx}
    x_l^\eta \approx x_l + J^{-1}(x_l) \eta.
\end{align}
Under the linearization regime, the $2\times 2$ covariance matrix of the reconstructed sinograms is given by
\begin{align}\label{SinoCovariance}
\textstyle \S_l = \S_{x_l} = J^{-1}(x_l) \S_{g^\eta}(x_l) J^{-t}(x_l),
\end{align}
where $-t$ denotes the transpose of inverse matrix (see eg. Cowan, p. 21)\cite{Cowan1998}. Here, the diagonal entries of \eqref{SinoCovariance} are the variances of the iodine and water sinograms, and the off-diagonal entries are the covariance between them (see, for example, Roessl et al. \cite{Roessl2007} for their explicit formulas).

We define the diagonal matrices iodine and water sinogram variances and covariance as $C^{(ij)}, i,j=1,2$. Therefore,
\begin{align}\label{sinovar}
    (C^{(ij)})_{ll} = (\S_l)_{ij}, \quad i,j=1,2.
\end{align}

 We also define a matrix $Q = (q_{ij})_{i,j=1, 2}$ with diagonal entries being the mean pixel variance of each material density map whereas the off-diagonal entries are the mean covariance between them. Let $B = (b_{pl})_{1\leq p\leq P,\; 1\leq l\leq L}$ denote the filtered back-projection matrix. Then,
\begin{align}
    q_{ij} = \frac{1}{P} {\rm{tr}}(BC^{(ij)}B^t), \quad j=1,2.
\end{align}

It is well known that the acquisition of photons for each spectral measurement may be shortened or lengthened to minimize reconstruction errors. Let $T$ be the total time of acquisition and $0< \tau < T$ be the acquisition time of the low energy measurement (so that $T-\tau$ is the time of acquisition in the high kV setting). Since the variance of the measurement noise is proportional to photon count, we observe that the covariance matrix of the measurement error is divided by $\tau$ when $i=1$ and $T-\tau$ with $i=2$. We thus have a modified covariance matrix for the reconstructed sinograms
\begin{align}\label{SinoCovarianceTime}
 \textstyle \S_l = J^{-1}(x_l) F\S_{g^\eta}(x_l) J^{-t}(x_l),
\end{align}
where 
$$F=
\begin{bmatrix}
\tau & 0\\
0 & T-\tau
\end{bmatrix}.
$$

It is important to note that the matrix $\S_l$ is always symmetric, and is also positive definite provided that $\det J(x_l) \neq 0$. We also observe from eqs. \eqref{SinoCovariance} and \eqref{SinoCovarianceTime} that the sinogram (co-)variances are inversely proportional to the square of the Jacobian determinant. Therefore, optimizing DECT scan parameters to maximize the minimum value of the Jacobian determinant over the sinogram domain leads to more stable sinogram reconstructions due to the reduced amplification of the measurement noise (also see Bal et al. \cite{BalGongTer}).

\subsection{Computation of pixel variances and the CNR of a monochromatic image}
Once the sinograms $x_1$ and $x_2$ (for iodine and water, respectively) are reconstructed from the DECT measurements using a nonlinear iterative algorithm, for which we use Newton's method in this study, they can be combined linearly to obtain a monochromatic sinogram, that is 
\begin{align}\label{xmono}
    x_{mono} = w_1x_1+w_2x_2 = w^tx,
\end{align}
where $w=[w_1,w_2]^t$ is a unit vector in $\R^2$, i.e, $\|w\|_2=1$.

 We note that for virtual monochromatic images (VMI), one considers the attenuation weights $w = \frac{M(E)}{\|M(E)\|_2}$ where the VMI energy $E$ is chosen to maximize a given metric, e.g. the contrast-to-noise ratio (CNR) of a region of interest (ROI) in a given imaging task.

 We define the contrast-to-noise ratio of a signal as the difference between the mean CT numbers of the signal and the background divided by the mean standard deviation of the signal:
\begin{align}\label{CNR}
    {\rm{CNR}}(w,tp,\tau) = \frac{mean(CT\#_{signal}) - mean(CT\#_{background})}{s(w,tp,\tau)_{signal}}.
\end{align}
Here, we consider as the background the part of the image containing only water.

In the following, we derive an analytic expression for the pixel variance $s^2(w,tp,\tau)$. 

Let $L$ denote the total number of lines (bins $\times$ views). If the monochromatic sinograms \eqref{xmono} corresponding to different lines are independent, then their covariance, which is denoted by $\Sigma_{mono}$ and is of size $L \times L$, is a diagonal matrix of monochromatic variances
\begin{align}\label{MonoSinoVarianceFixed}
\sigma^2_l(w,tp,\tau) = w^t \S_lw.
\end{align}

The pixel covariance matrix is then given by
\begin{align}
 \S_y &= B \Sigma_{mono} B^t,
\end{align}
where $B = (b_{pl})_{1\leq p\leq P,\; 1\leq l\leq L}$ is the filtered back-projection matrix. 

By direct calculation, we obtain that
\begin{align}\label{VMIvariance}
 (\S_y)_{ii} =\sum_{l=1}^L \sigma^2_l(w,tp,\tau)b_{pl}^2, \quad i=1,\dots, P.
\end{align}

Therefore, the mean pixel variance of the monochromatic signal is given by
\begin{align}\label{VMIvar_ave}
s^2(w,tp,\tau) &= \frac{1}{P}\sum_{p=1}^P (\S_y)_{ii} 
 = \frac{1}{P} \sum_{p=1}^P \sum_{l=1}^L \sigma^2_l(w,tp,\tau)b_{pl}^2 \nonumber \\
 &= \frac{1}{P}\sum_{l=1}^L \big({\textstyle \sum}_{p=1}^P b_{pl}^2\big) \sigma^2_l(w,tp,\tau) 
 = \frac{1}{P}\sum_{l=1}^L(b_l^tb_l)\sigma^2_l(w,tp,\tau),
\end{align}
where $b_l$ denote the $l$-th column of $B$. Now using eq. \eqref{MonoSinoVarianceFixed}, we obtain that
\begin{align}\label{VMIvar_ave2}
s^2(w,tp,\tau) 
& = \frac{1}{P} \sum_{l=1}^L(b_l^tb_l) w^t \S_l w 
=  w^t \left(\frac{1}{P}\sum_{l=1}^Lb_l^tb_l \S_l \right) w =  w^t Q w, 
\end{align}
since
\begin{align}
    q_{ij} = \frac{1}{P} {\rm{tr}(BC^{(ij)}B^t)} = \frac{1}{P} \sum_{l=1}^L  \sum_{p=1}^P b_{pl}^2 (\S_l)_{ij} = \frac{1}{P} \sum_{l=1}^Lb_l^tb_l (\S_l)_{ij}  , \quad j=1,2.
\end{align}

\subsection{Minimization of mean pixel variance} \label{sec:maxCNR}
In the following, we show that the smallest eigenvalue of the mean pixel covariance matrix gives the minimum mean variance for VMI obtained by using optimal VMI energy, identifying optimal energy levels for maximum image quality.

We first observe that if $\det J(x_l) \neq 0$ for all $l$, then $\S_l$ is positive definite for all $l$, and thus the matrix $Q=\textstyle \frac{1}{P} \sum_{l=1}^Lb_l^tb_l \S_l$ is positive definite. This implies that $s^2(w,tp,\tau)=w^t Q w$ is a positive definite quadratic form.

When the matrix $Q$ is positive definite, by the spectral theorem, it is orthogonally diagonalizable, that is the eigenvectors of $Q$ form an orthogonal basis for $\R^2$.
For $i=1,2$, we let $Qu_i=\lambda_i u_i$ where $\|u_i\|_2=1$, and $\lambda_1 \geq \lambda_2 > 0$, i.e., $u_i$ is the unit eigenvector associated to eigenvalue $\lambda_i$ (see eg. Horn and Johnson \cite{Horn}). Then, for each tube potential pair $tp$ and fluence $\tau$,
\begin{align}\label{min_w}
\min_{\|w\|_2=1} s^2(w,tp,\tau) = \min_{\|w\|_2=1} w^t Q(tp,\tau) w = \lambda_2(tp,\tau), 
\end{align}
where the minimizer is $u_2(tp,\tau)$. One can then minimize $\lambda_2$ over $tp$ and $\tau$ to find the optimal tube potential pair and the photon fluence. Hence, 
\begin{align}\label{thm1}
(w^*,tp^*,\tau^*)= \argmin_{w,tp,\tau} s^2(w,tp,\tau),
\end{align}
where 
\begin{align}
    (tp^*,\tau^*)= \argmin_{tp,\tau} \lambda_2(tp,\tau),
\end{align}
and $w^*$ is the unit eigenvector of $Q(tp^*,\tau^*)$ corresponding to $\lambda_2(tp^*,\tau^*)$.

We observe that if the Jacobian determinant values increase, then the diagonal entries of ($\S_l$ and hence) $Q$ decrease. Since the smallest eigenvalue of a $2\times2$ matrix is always less than the smallest diagonal entry (see eg. Horn and Johnson \cite{Horn}), scan parameters that maximize Jacobian determinant values give the minimal smallest eigenvalue of $Q$ and hence it reduces the mean pixel variance.

In general, when the mean pixel variance is minimized only over the set of the attenuation weights, one expects to obtain a larger value in comparison to generalized weights. However, for the iodine-water material pair, we numerically observed that there is $E^*$ such that $\frac{M(E^*)}{\|M(E^*)\|_2} = w^*$. This is mainly because $M_1(E)>M_2(E)$ for all $E$ in the diagnostic energy range and the components of $w^*$ have the same sign. We thus have
\begin{align}\label{min_wE}
\min_{0<E<E_{max}} s^2(E,tp,\tau) = \min_{\|w\|_2=1} s^2(w,tp,\tau) = \lambda_2(tp,\tau).
\end{align} 
We remark that this result is specific to iodine-water material pair and may not hold for others.

\section{Results}\label{sec:results}
In this section, we present the results of our numerical experiments conducted for the iodine-water material pair. We first explain our experimental setup. We then demonstrate the dependence of uniqueness to the Jacobian determinant values and showcase a non-unique reconstruction. We finally provide our numerical results on the optimization of the VMI energy and iodine CNR in virtual monochromatic images.

\subsection{Experimental setup}\label{sec:experiment setup}
\begin{figure}[H]
\centering
\includegraphics[width=0.9\textwidth]{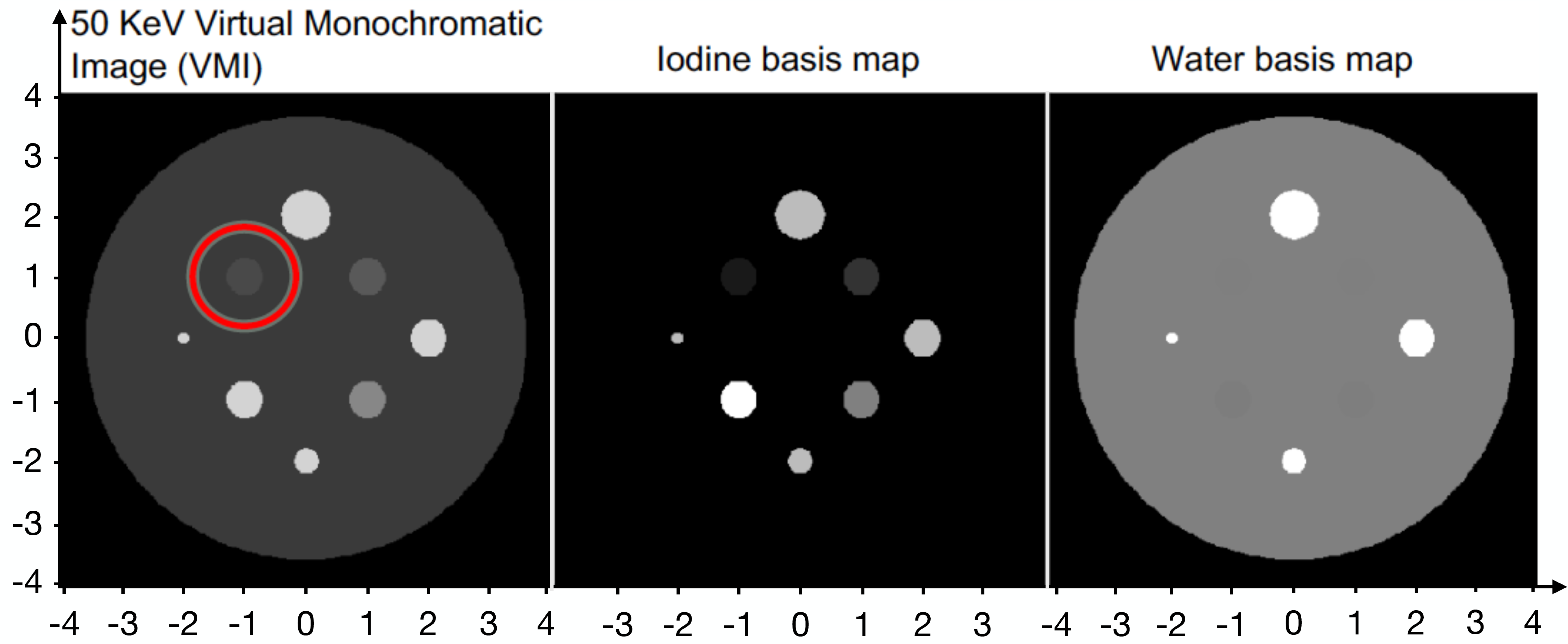}
\caption{The plot of the circular water phantom with inserted iodine-solution and calcium disk signals. The red circle indicates the disk with the lowest iodine concentration, which is our region of interest.}
\label{fig:phantom}
\end{figure}

In the numerical experiments, we used a $4\times4\; cm^2$ phantom consisting of a large water disk with inserted iodine-solution and calcium disk signals (see Fig. \ref{fig:phantom}). The centers and radii of each disk, and the concentration of iodine solutions are given in table \ref{t:phantom}.

\vspace{2em}
\begin{table}[htbp]
\centering
\begin{tabular}{|c|c|c|}
\hline
\multicolumn{1}{|c|}{Material} & Center (cm) & Radius (cm)   \\ \hline
0.1\% Iodine                   & (-1,1)      & 0.3           \\ \hline
0.2\% Iodine                   & (1,1)       & 0.3           \\ \hline
0.5\% Iodine                   & (1,-1)      & 0.3           \\ \hline
1\% Iodine                     & (-1,1)      & 0.3           \\ \hline
Calcium                        & (-2,0)      & 0.1           \\ \hline
Calcium                        & (0,-2)      & 0.2           \\ \hline
Calcium                        & (2,0)       & 0.3           \\ \hline
Calcium                        & (0,2)       & 0.4           \\ \hline
Water                          & (0,0)       & 3.6           \\ \hline
\end{tabular}
\caption{Phantom configuration.}
\label{t:phantom}
\end{table}

For the phantom shown in Fig. \ref{fig:phantom}, the DECT measurements were numerically simulated for $512 \times 512$ lines (bins $\times$ views) in fan beam geometry. The distance from source to detector was 100 cm. For each line, 100 realizations were obtained by considering a compound Poisson process\cite{Whiting2002}. The range of x-ray tube voltages was considered to be 30-150 kV. The x-ray beam filter material was fixed at 2mm of Aluminum. The results were not significantly affected by variations in the total number of photons and fluence between low and high kV scans. As a result, these factors were considered equal in the analysis.

\subsection{Uniqueness of the reconstructed sinograms}\label{sec:uniqueness}
As mentioned in section \ref{subsec:Jacobian}, the uniqueness of reconstructions is ensured if the Jacobian determinant is nonzero everywhere in the rectangle containing all possible sinogram values (or pathlengths). Jacobian determinant values vary according to the chosen tube potentials and the filters of the x-ray energy spectra, where we fix the latter and focus on examining the effect of the former. 

\begin{figure}[H]
\centering
\includegraphics[width=0.6\textwidth]{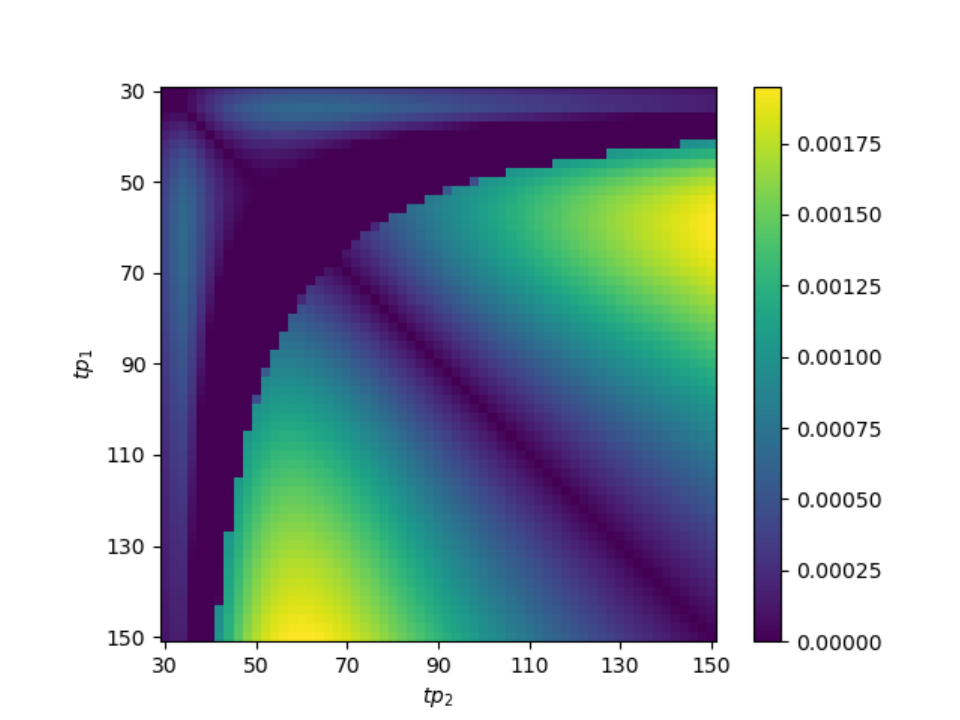}
\caption{The minimum of absolute value of Jacobian determinants over pathlengths up to 0.01 and 7.2, for iodine and water respectively, as a function of tube settings varying from 30 to 150 kV.}
\label{fig:signedDetJ}
\end{figure}

According to the eqs. \eqref{SinoCovariance} and \eqref{SinoCovarianceTime}, the sinogram (co-)variances are inversely proportional to the square of the Jacobian determinant. In Figure \ref{fig:signedDetJ}, we present the plot of the quantity
\begin{align}\label{minAbsDetJ}
\min_{x \in \mathbb{R}} |\det J(x)|,
\end{align}
This is the minimum of absolute values of Jacobian determinant over the rectangular region $\mathcal{R} = [0,0.01] \times [0,7.2]$, which corresponds to all possible pathlengths for iodine and water obtained from our phantom (depicted in Figure \ref{fig:phantom}). We caution that the Jacobian determinant also depends on the tube settings, although we omit the specific notation for simplicity.

This plot reveals that lighter regions, which correspond to higher absolute values of the Jacobian determinant, indicate tube potentials that lead to more stable transmission-to-sinogram transformation. Therefore, the use of the tube potentials in the region $50 \le tp_1 \le 80$ and $120 \le tp_2 \le 150$ yield more stable sinogram reconstructions.

\begin{figure}[H]
\centering
\includegraphics[width=0.5\textwidth]{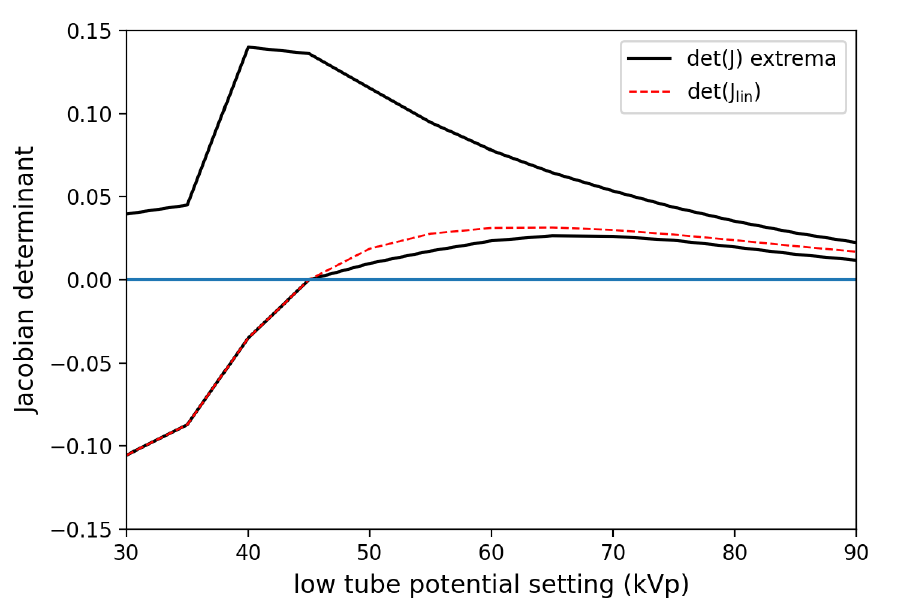}
\caption{Plot of the Jacobian determinant for the
map that transforms iodine/water sinograms to
dual-energy transmission at low and high kV. The
black curves show the range of determinant values
over the expected tissue path lengths and the red
curve shows the determinant for the linearized
function. For the shown results, the high kV setting
is 120, and the low kV setting is on the x-axis.
Maximizing determinant values leads to stable
transmission-to-sinogram transformation.}
\label{fig:detJplot}
\end{figure}

In Fig. \ref{fig:detJplot}, for a fixed high tube setting of $tp_2=120$ kV, we plot in black the extremal Jacobian determinant values attained in the rectangle $ \rR = [0,0.01]\times[0,7.2]$ cm$^2$ for low tube setting varying from 30 to 90 kV. The red curve represents the Jacobian determinant of the linearized map, which also corresponds to the Jacobian determinant values at zero pathlength. 

\begin{figure}[H]
\centering
\includegraphics[width=0.5\textwidth]{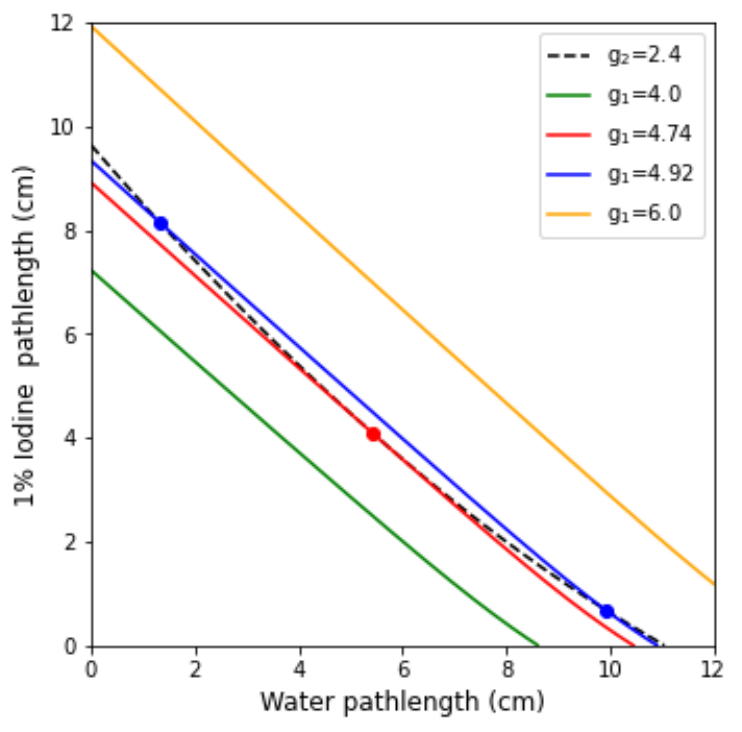}
\caption{Plot of the level curves of the log-intensity measurements as a function of the water and iodine sinogram values demonstrating the existence of singular solutions to the inverse sinograms-to-DECT measurements map. The red dot is a point where the Jacobian vanishes. The dashed black curve ($g_2=2.4$) intersects the blue one ($g_1=4.92$) at two twice (shown with blue dots), so both sinogram value pairs lead to the same log-intensity measurement value of $(g_1,g_2)=(4.92,2.4)$.}
\label{fig:nonUniqueness}
\end{figure}

We observe in Fig. \ref{fig:detJplot} that if the low tube potential is less than 45 kV, the Jacobian determinant vanishes somewhere inside the rectangle, and hence the uniqueness of sinogram reconstructions is not guaranteed. In fact, for the tube potential pair $(tp_1,tp_2)=(35,120)$, for which the Jacobian determinant becomes zero near the sinogram values $x = (x_1,x_2)=(4,6)$, we encountered singular solutions to the inverse sinograms-to-DECT measurements map. In Fig. \ref{fig:nonUniqueness}, we present a plot of the level curves of the corresponding log-intensity measurements as a function of the iodine and water sinogram values. The red dot is where the Jacobian determinant is zero.  The level curve of $g_2=2.4$ (shown in dashed black) intersects the level curve of $g_1=4.92$ (shown in blue) at two points (shown with blue dots). This means that two different sinogram value pairs (shown with blue dots) lead to the same measurement value of $(g_1,g_2)=(4.92,2.4)$.
\subsection{Optimization of the iodine CNR in virtual monochromatic images}
In this section, we present the results of the analysis conducted on the Contrast-to-Noise Ratio (CNR) values of the 0.1\% Iodine disk in the virtual monochromatic image (VMI) which is enclosed by the red circle in Fig. \ref{fig:phantom}. To compare the analytical and empirical approaches, the CNR values were computed using equation \eqref{CNR} analytically and through image reconstruction for 100 realizations. In the calculations, the 0.1\% Iodine disk and the water only parts of the phantom were considered as the signal and the background, respectively. 

\begin{figure}[H]
\centering
\includegraphics[width=0.5\textwidth]{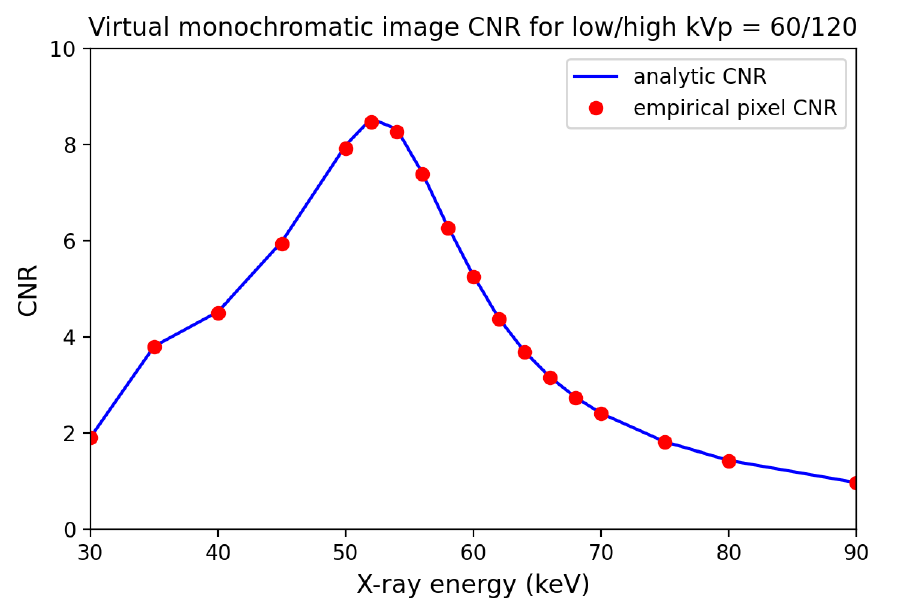}
 
\caption{For low/high tube settings of 60/120 kV,
the CNR of the 0.1\% Iodine disk is shown for a noise
level of 50,000 photons incident on each detector
pixel. Shown are the analytically computed CNR
values with blue curve and the empirically calculated
values using noise realizations, with the red dots.
The agreement between theory and simulation
validates the computational method. A clear peak in
the 0.1\% Iodine disk CNR is seen in the VMI for an
energy of 55 keV.}
\label{fig:VMI-CNRvsE}
\end{figure}

Fig. \ref{fig:VMI-CNRvsE} displays these CNR values for the low and high tube settings of 60/120 kV, with the blue curve representing the analytically computed values and the red dots indicating the empirically obtained results. Notably, there is a remarkable agreement between the empirical and analytical findings. Furthermore, the maximum CNR for the 0.1\% Iodine disk in the Virtual Monochromatic Images (VMI) was observed at an energy level of 55 keV. Similar results were obtained for the 60/140 kV setting. These findings provide valuable insights into the optimal energy range for maximizing CNR in VMI. 

\begin{figure}[H]
\centering
\includegraphics[width=0.5\textwidth]{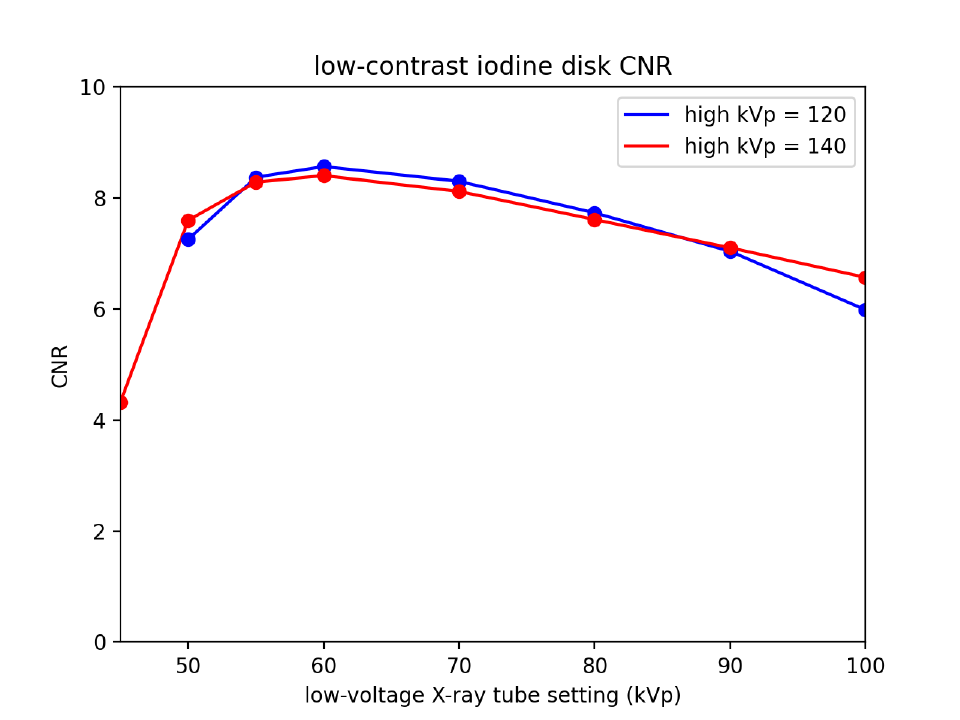}
 
\caption{Using the VMI energy that maximizes the
0.1\% Iodine disk CNR, the low and high kV settings are
varied. Shown in the plot are only the high kV
settings of 120 and 140; the low kV setting is
indicated on the x-axis. Of the values computed, the
60/120 kV setting allows for maximum 0.1\% Iodine CNR.}
\label{fig:iodineCNR}
\end{figure}

To further explore the influence of tube potentials on the CNR, we plot in Fig. \ref{fig:iodineCNR} the CNR values for various tube settings at this optimal VMI energy of 55 keV. Comparing the high tube potential settings of 120 kV and 140 kV, it is observed that the former yields slightly higher CNR values. Consequently, the tube potential pair of 60/120 kV emerges as the optimal choice for maximizing the CNR of the 0.1\% Iodine disk in VMI within the scope of this study. We remark that, in view of Fig. \ref{fig:signedDetJ}, this pair of tube potentials lead to stable transmission-to-sinogram transformation. This result highlights the significance of selecting the appropriate kV settings to optimize the visualization and diagnostic quality of the Iodine disk in VMI imaging. 

\section{Discussion and conclusion}\label{sec:conclusion}
In this study, we have introduced a novel approach for optimizing the DECT scan protocol specifically for iodine-based CNR. Our methodology centers around analyzing the propagation of noise from DECT measurements to material density images. By linearizing the inverse sinogram-to-transmission measurements map, we were able to derive analytic expressions for the mean sinogram and pixel variances. These expressions serve as key components of our framework, enabling us to systematically optimize the DECT scan parameters for improved image quality in terms of iodine-based CNR. We also identified the ideal VMI energy that maximizes CNR. We have shown that the VMI synthesized from iodine and water density images with the optimal energy exhibits the least amount of noise. As a result, our findings indicate that employing more general weights to linearly combine these images does not provide any substantial improvement in image quality.

Our DECT simulations were based on dual source DECT scanners. However, our theoretical framework is applicable to dual-layer detectors and photon counting detectors as well. In those cases, one needs to optimize the energy window thresholds instead of the tube potentials. The photon fluence was taken into account in our framework. However, we considered equal times for both energy measurement in the numerical simulations as its influence on the results was not significant. We remark that the results of this study are specific to the phantom used as it directly affects the size of the sinogram domain and consequently the range of Jacobian values observed. Our primary objective in this research was to propose a framework for optimizing DECT scan parameters to achieve maximum stability of reconstructions and contrast-to-noise ratio (CNR). It is important to consider the specific imaging task at hand and carefully assess all relevant factors when implementing this framework.

In a future study, we plan to test our theoretical results using real DECT data. We also plan to extend our analysis to the case of three or more materials and energy measurements.
\section*{Acknowledgment}
F. Terzioglu's work was supported in part by the NSF DMS grant 2206279. G. Bal's work was supported in part by the NSF DMS grants 2306411 and 1908736. This work is also supported in part by NIH Grant Nos. R01-EB023968 and R21-CA263660. The contents of this article are solely the responsibility of the authors and do not necessarily represent the official views of the National Institutes of Health.

\bibliographystyle{medphy}
\bibliography{references}

\begin{thebibliography}{10}

\bibitem{Grant2014}
K.~L. Grant, T.~G. Flohr, B.~Krauss, M.~Sedlmair, C.~Thomas, and B.~Schmidt,
\newblock Assessment of an advanced image-based technique to calculate virtual
  monoenergetic computed tomographic images from a dual-energy examination to
  improve contrast-to-noise ratio in examinations using iodinated contrast
  media,
\newblock Investigative radiology {\bf 49}, 586--592 (2014).

\bibitem{Leng2015}
S.~Leng, L.~Yu, J.~G. Fletcher, and C.~H. McCollough,
\newblock Maximizing iodine contrast-to-noise ratios in abdominal CT imaging
  through use of energy domain noise reduction and virtual monoenergetic
  dual-energy CT,
\newblock Radiology {\bf 276}, 562 (2015).

\bibitem{Kalisz2018}
K.~Kalisz, N.~Rassouli, A.~Dhanantwari, D.~Jordan, and P.~Rajiah,
\newblock Noise characteristics of virtual monoenergetic images from a novel
  detector-based spectral CT scanner,
\newblock European Journal of Radiology {\bf 98}, 118--125 (2018).

\bibitem{Sakabe2018}
D.~Sakabe, Y.~Funama, K.~Taguchi, T.~Nakaura, D.~Utsunomiya, S.~Oda, M.~Kidoh,
  Y.~Nagayama, and Y.~Yamashita,
\newblock Image quality characteristics for virtual monoenergetic images using
  dual-layer spectral detector CT: comparison with conventional tube-voltage
  images,
\newblock Physica Medica {\bf 49}, 5--10 (2018).

\bibitem{Tabari2020}
A.~Tabari, M.~S. Gee, R.~Singh, R.~Lim, K.~Nimkin, A.~Primak, B.~Schmidt, and
  M.~K. Kalra,
\newblock Reducing radiation dose and contrast medium volume with application
  of dual-energy CT in children and young adults,
\newblock American Journal of Roentgenology {\bf 214}, 1199--1205 (2020).

\bibitem{Hounsfield}
G.~N. Hounsfield,
\newblock Computerized transverse axial scanning (tomography): {P}art 1.
  Description of system,
\newblock The British Journal of Radiology {\bf 46}, 1016--1022 (1973).

\bibitem{AlvarezMacovski}
R.~E. Alvarez and A.~Macovski,
\newblock Energy-selective reconstructions in x-ray computerised tomography,
\newblock Physics in Medicine \& Biology {\bf 21}, 733 (1976).

\bibitem{Levine}
Z.~H. Levine,
\newblock Nonuniqueness in dual-energy {CT},
\newblock Medical physics {\bf 44}, e202--e206 (2017).

\bibitem{Alvarez2019}
R.~E. Alvarez,
\newblock Invertibility of the dual energy x-ray data transform,
\newblock Medical Physics {\bf 46}, 93--103 (2019).

\bibitem{BalTer20}
G.~Bal and F.~Terzioglu,
\newblock Uniqueness criteria in multi-energy {CT},
\newblock Inverse Problems {\bf 36}, 065006 (2020).

\bibitem{BalGongTer}
G.~Bal, R.~Gong, and F.~Terzioglu,
\newblock An inversion algorithm for P-functions with applications to
  multi-energy CT,
\newblock Inverse Problems {\bf 38}, 035011 (2022).

\bibitem{Tao2019}
S.~Tao, K.~Rajendran, W.~Zhou, J.~G. Fletcher, C.~H. McCollough, and S.~Leng,
\newblock Improving iodine contrast to noise ratio using virtual monoenergetic
  imaging and prior-knowledge-aware iterative denoising (mono-PKAID),
\newblock Physics in Medicine \& Biology {\bf 64}, 105014 (2019).

\bibitem{Yu2009}
L.~Yu, A.~N. Primak, X.~Liu, and C.~H. McCollough,
\newblock Image quality optimization and evaluation of linearly mixed images in
  dual-source, dual-energy CT,
\newblock Medical physics {\bf 36}, 1019--1024 (2009).

\bibitem{Yu2011}
L.~Yu, J.~A. Christner, S.~Leng, J.~Wang, J.~G. Fletcher, and C.~H. McCollough,
\newblock Virtual monochromatic imaging in dual-source dual-energy CT:
  radiation dose and image quality,
\newblock Medical physics {\bf 38}, 6371--6379 (2011).

\bibitem{Michalak2017}
G.~Michalak, J.~Grimes, J.~Fletcher, A.~Halaweish, L.~Yu, S.~Leng, and
  C.~McCollough,
\newblock Selection of optimal tube potential settings for dual-energy CT
  virtual mono-energetic imaging of iodine in the abdomen,
\newblock Abdominal Radiology {\bf 42}, 2289--2296 (2017).

\bibitem{Dabli2021}
D.~Dabli, J.~Frandon, A.~Hamard, A.~Belaouni, T.~Addala, J.-P. Beregi, and
  J.~Greffier,
\newblock Optimization of image quality and accuracy of low iodine
  concentration quantification as function of kVp pairs for abdominal imaging
  using dual-source CT: A phantom study,
\newblock Physica Medica {\bf 88}, 285--292 (2021).

\bibitem{Ren2021}
Y.~Ren, H.~Xie, W.~Long, X.~Yang, and X.~Tang,
\newblock On the Conditioning of Spectral Channelization (Energy Binning) and
  Its Impact on Multi-Material Decomposition Based Spectral Imaging in
  Photon-Counting CT,
\newblock IEEE Transactions on Biomedical Engineering {\bf 68}, 2678--2688
  (2021).

\bibitem{Williamson2006}
J.~F. Williamson, S.~Li, S.~Devic, B.~R. Whiting, and F.~A. Lerma,
\newblock On two-parameter models of photon cross sections: application to
  dual-energy CT imaging,
\newblock Medical physics {\bf 33}, 4115--4129 (2006).

\bibitem{Whiting2002}
B.~R. Whiting,
\newblock Signal statistics in x-ray computed tomography,
\newblock in {\em Medical Imaging 2002: Physics of Medical Imaging}, volume
  4682, pages 53--60, International Society for Optics and Photonics, 2002.

\bibitem{NIST}
J.~H. Hubbell and S.~M. Seltzer,
\newblock Tables of X-ray mass attenuation coefficients and mass
  energy-absorption coefficients 1 ke{V} to 20 {M}e{V} for elements {Z}= 1 to
  92 and 48 additional substances of dosimetric interest,
\newblock Technical report, National Inst. of Standards and Technology-PL,
  Gaithersburg, MD (United~States), 1995.

\bibitem{Spekpy2}
G.~Poludniowski, A.~Omar, R.~Bujila, and P.~Andreo,
\newblock SpekPy v2. 0—a software toolkit for modeling x-ray tube spectra,
\newblock Medical Physics {\bf 48}, 3630--3637 (2021).

\bibitem{GaleNikaido}
D.~Gale and H.~Nikaido,
\newblock The {J}acobian matrix and global univalence of mappings,
\newblock Mathematische Annalen {\bf 159}, 81--93 (1965).

\bibitem{Garcia}
C.~Garcia and W.~Zangwill,
\newblock On univalence and {P}-matrices, Report 7737,
\newblock Center for Mathematical Studies in Business and Economics, University
  of Chicago  (1977).

\bibitem{MasColell}
A.~Mas-Colell,
\newblock Homeomorphisms of Compact, Convex Sets and the {J}acobian Matrix,
\newblock SIAM Journal on Mathematical Analysis {\bf 10}, 1105--1109 (1979).

\bibitem{Cowan1998}
G.~Cowan,
\newblock {\em Statistical Data Analysis},
\newblock Oxford University Press, 1998.

\bibitem{Roessl2007}
E.~Roessl, A.~Ziegler, and R.~Proksa,
\newblock On the influence of noise correlations in measurement data on basis
  image noise in dual-energylike x-ray imaging,
\newblock Medical physics {\bf 34}, 959--966 (2007).

\bibitem{Horn}
R.~A. Horn and C.~R. Johnson,
\newblock {\em Matrix Analysis},
\newblock Cambridge University Press, 2012.

\end{thebibliography}

\end{document}